\documentclass[aps,prb,a4paper,twocolumn,superscriptaddress,floatfix,showpacs,preprintnumbers,amsmath,amssymb]{revtex4}
\usepackage{url}
\usepackage[english]{babel}
\usepackage[utf8]{inputenc}
\usepackage{amsmath}
\usepackage{graphicx,epstopdf}
\usepackage{amssymb,epsfig,color,textcase}
\usepackage{hyperref}
\usepackage{doi}
\providecommand{\U}[1]{\protect\rule{.1in}{.1in}}

\begin{document}


\title{Orbital-selective behavior in cubanite CuFe$_2$S$_3$}

\author{Elizaveta A. Pankrushina}
\affiliation{A.N. Zavaritsky Institute of Geology and Geochemistry, Ural Branch of the Russian Academy of Sciences, Ak. Vonsovskogo St. 15, 620016, Ekaterinburg, Russia}

\author{Alexey V. Ushakov}%
\affiliation{M.N. Miheev Institute of Metal Physics of Ural Branch of Russian Academy of Sciences, S. Kovalevskaya St. 18, 620990 Ekaterinburg, Russia}

\author{Mohsen M. Abd-Elmeguid}
\affiliation{II. Physikalisches Institut, Universit\"at zu Koln, Z\"ulpicher Strasse 77, 50937 K\"oln, Germany}

\author{Sergey V. Streltsov}%
\email{streltsov@imp.uran.ru}
\affiliation{M.N. Miheev Institute of Metal Physics of Ural Branch of Russian Academy of Sciences, S. Kovalevskaya St. 18, 620990 Ekaterinburg, Russia}
\affiliation{Department of theoretical physics and applied mathematics, Ural Federal University, Mira St. 19, 620002 Ekaterinburg, Russia}

\date{\today}

\begin{abstract}
Using {\it ab initio} band structure calculations we show that mineral cubanite, CuFe$_2$S$_3$, demonstrates an orbital-selective behavior with some of the electrons occupying molecular orbitals of $x^2-y^2$ symmetry and others localized at atomic orbitals.  This is a rare situation for $3d$ transition metal compounds explains experimentally observed absence of charge disproportionation, anomalous M\"ossbauer data, and ferromagnetic ordering in between nearest neighbor Fe ions. 
\end{abstract}


\maketitle

\section {Introduction \label{Intro}}

In recent years there has been growing interest in investigating the ground state properties of transition metal (TM) compounds in which the orbital degrees of freedom play a dominant role~\cite{Jackeli2009,Liu2018,Feiguin2019a,Streltsov2020,Li2021}. Particularly interesting is the directional character of the $d$ orbitals that can lead to the formation of various electronic and magnetic ground states, see e.g. Ref.~[\citenum{Khomskii2020}]. Indeed, pronounced effects have been recently reported for many $4d$ and $5d$ metal cluster compounds revealing that different orbitals can behave in different ways, i.e. demonstrate orbital-selective behavior: orbitals directed toward neighbors in a TM dimer (or trimer) behave as delocalized and can be described by molecular orbitals, whereas the electrons in other orbitals are localized. This can result in a suppression of the effective magnetic moment and strongly affects the mechanism of exchange interaction. Thus, it turns out that the ground state properties of such cluster compounds are determined by their orbital structure and the related orbital-selective behavior.  

However, in general such a situation is more typical for $4d-5d$ systems, since a direct overlap between more spatially extended $4d-5d$ orbitals is much larger than in case of $3d$. In this respect, ternary chalcogenide compound CuFe$_2$S$_3$ (mineral cubanite) is a promising candidate for such study, since its electronic and magnetic ground state properties are far from being clear. CuFe$_2$S$_3$ crystallizes in the orthorhombic structure with a single class of Fe\cite{Szymanski-74,Fleet-70,Goh-10,Buerger-45,Buerger-47}. It was argued that iron has a nominal valency of +2.5, due to a ``rapid electron exchange'' between Fe$^{2+}$ and Fe$^{3+}$ ions in Fe--Fe dimers\cite{Imbert-67,Greenwood-68}. The Fe$^{2+}$–Fe$^{3+}$ ion pairs are tetrahedrally coordinated by S atoms with two FeS$_4$ tetrahedra sharing common edges, forming a cluster of paired FeS$_4$ tetrahedra. More structural details and given below in Section~\ref{structure}. We note that $^{57}$Fe M\"ossbauer measurements also indicate that there is only one Fe site and the data obtained are explained assuming ``rapid electron exchange'' between Fe$^{2+}$ ($3d^6$) and Fe$^{2+}$($3d^5$), giving an intermediate valency of about 2.5. Such fast electronic fluctuation would imply metallic conductivity, however, resistivity measurement on natural single crystal of CuFe$_2$S$_3$ have shown n-type semiconductor behavior\cite{Sleight-73}.

Magnetic measurements on CuFe$_2$S$_3$ show that cubanite is canted antiferromagnet (AFM) at low-temperature phase. Interestingly enough nearest iron neighbors inside a Fe--Fe dimer are coupled not antiferromagnetically as a conventional superexchange theory would suggest (see e.g.  Ref.~[\citenum{Khomskii2020}] or Ref.~[\citenum{Goodenough}]), but ferromagnetically. The Fe--Fe dimers are coupled to each other antiferromagnetically~\cite{Wintenberger1974,Townsend-73}. The local magnetic moment was found to be $\sim$3.2$\mu_B$, i.e. somewhat smaller than what one would expect even for Fe$^{2+}$.~\cite{Wintenberger1974}

\begin{figure}[t]
\includegraphics[width=0.4\textwidth]{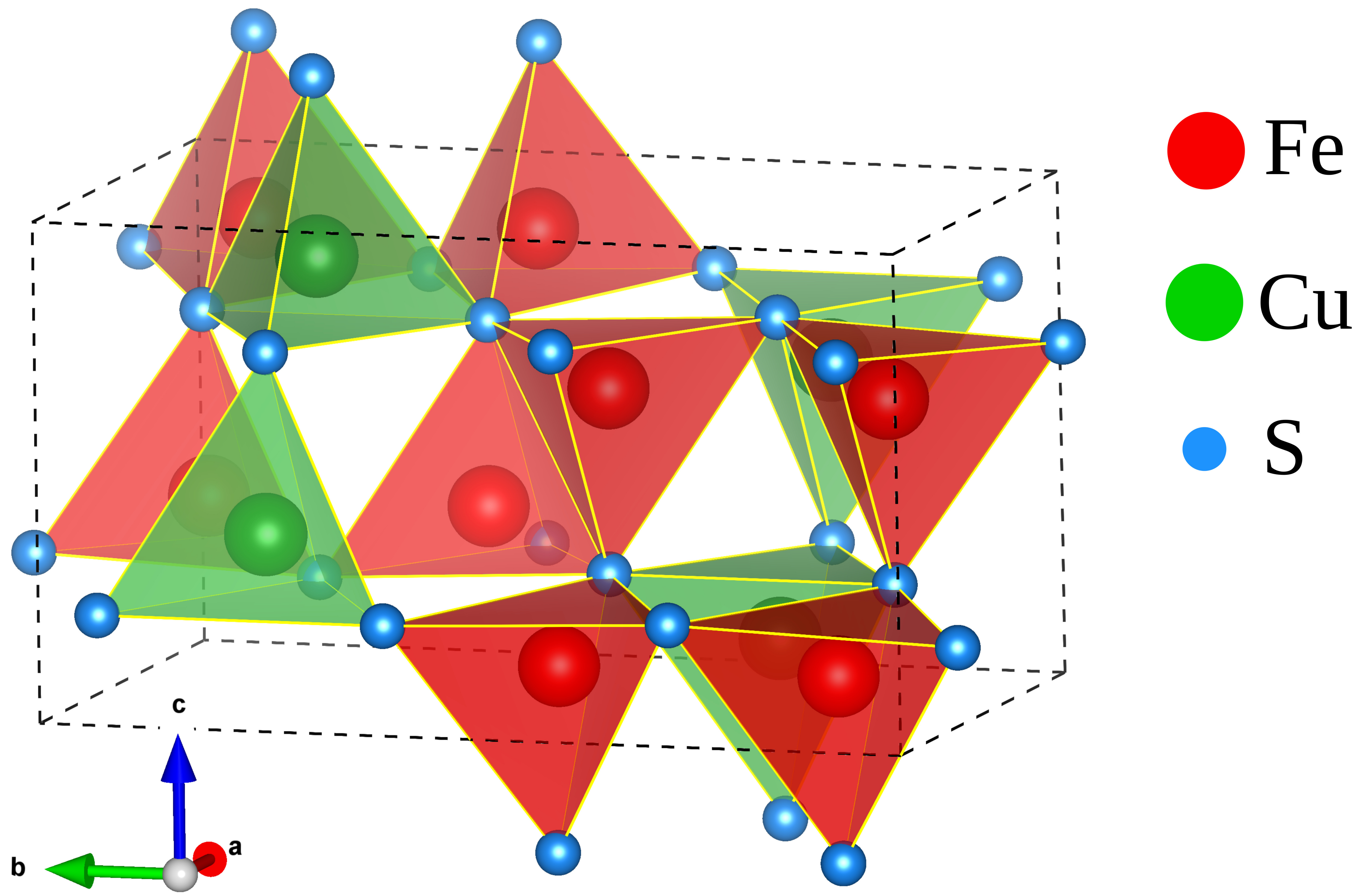}
\caption{\label{str} The crystal structure of CuFe$_2$S$_3$. FeS$_4$ (CuS$_4$) tetrahedra are shown in red (green) colour. Two neighboring FeS$_4$ tetrahedra have a common edge.}
\end{figure}

Motivated by the above mentioned physical aspects and in view of the current research activity on CuFe$_2$S$_3$ for several technological applications, we have investigated the ground state of CuFe$_2$S$_3$ using GGA and GGA+U calculations. We were able to unveil the ground state properties of CuFe$_2$S$_3$ and show that there are two types of $d$ orbitals in cubanite. First ones form molecular orbitals for two neighboring irons and electrons occupying these orbitals belong to both ions. Second are localized at atomic sites. This not only explains an intermediate valence 2.5+ of Fe and anomalous results of M\"ossbauer measurements, but also elucidates the origin of the unexpected magnetic structure of this compound.

\section {Crystal structure \label{structure}}

The cubanite CuFe$_2$S$_3$ crystallizes within an orthorhombic structure (space group $Pnma$, a = 6.23~\AA,~b = 11.11~\AA~and c = 6.46~\AA) with a quasi-hexagonal stacking of S$^{2-}$ anions where the cations are tetrahedrally coordinated (see Fig. \ref{str})\cite{Fleet-70}. The structure is based upon a hexagonal close-packed network of S atoms with the cations in ordered, tetrahedral sites; the Cu atoms and 1/3 of the S atoms occupy the equipoint 4c (mirror planes), and the Fe atoms and remaining 2/3 of the S atoms are in the general positions, 8d ~\cite{Fleet-70}. Experimental atomic positions and lattice parameters are taken from Ref.~[\citenum{Buerger-47}].

In orthorhombic cubanite Cu and Fe atoms are tetrahedrally coordinated by S atoms with two FeS$_4$ tetrahedra sharing their edges. This gives pairs of Fe ions with rather short distances between them~\cite{Szymanski-74,Fleet-70,Goh-10,Buerger-45,Buerger-47}. distance of 2.81 Å, however, is too long to represent a chemical bond. It is also larger than that in KFeS$_2$ where tetrahedrally-coordinated iron atoms form chains with the Fe--Fe distance of 2.7 Å~\cite{Hullige-83}.

\section {Calculation details}
\begin{figure}[t]
\includegraphics[width=0.4\textwidth]{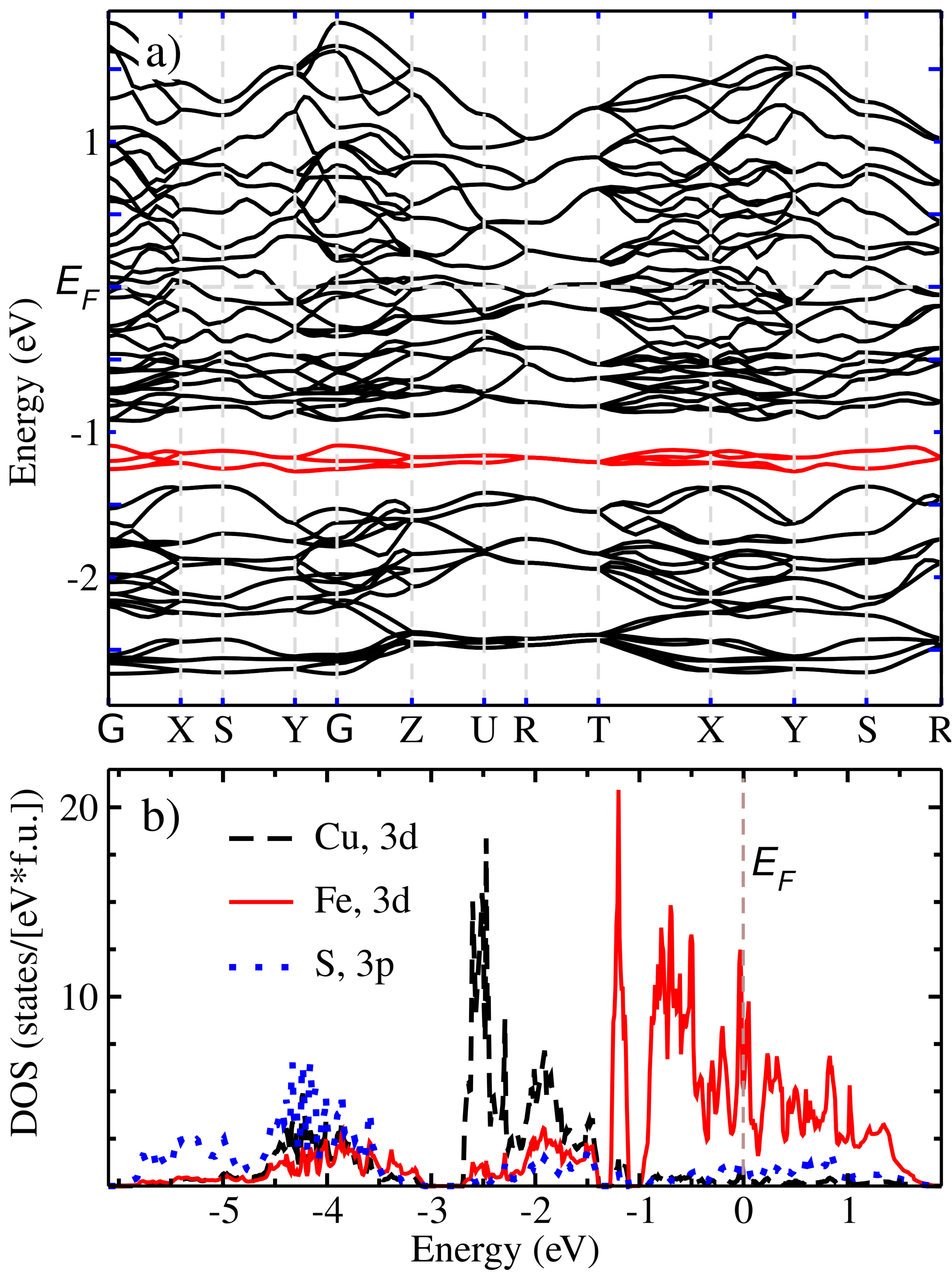}
\caption{\label{gga:dos} The band structure (a) and the partial densities of states (b) of CuFe$_2$S$_3$ obtained in non-magnetic GGA calculations. The Fermi level is in zero. One may note 4 isolated Fe-$3d$ bands 
between $-1.1$ and $-1.6$ eV (marked in red lines), which correspond to two bonding molecular states.}
\end{figure}

The {\it ab initio} band structure calculations of CuFe$_2$S$_3$ were carried out within the framework of density functional theory (DFT)~\cite{Hohenberg-64} implemented in VASP package~\cite{Kresse-96}. We used the generalized gradient approximation (GGA)~\cite{Perdew-96} and projector augmented wave (PAW) method~\cite{Bloechl-94}. The exchange-correlation functional in Perdew-Burker-Ernzerhof (PBE) form was utilized~\cite{Perdew-97}. The cut-off energy was chosen to be 600~eV and the mesh of $6\times3\times6$ was used for integration over the Brillouin zone according to Monkhorst-Pack scheme~\cite{Monkhorst-76}. A non-interacting GGA Hamiltonian for the estimation of hopping integrals inside the Fe $3d$ states was generated using the Wannier projection procedure~\cite{Korotin-08} in Quantum Espresso code~\cite{Gianozzi-09} on the same k-point grid.
The correlation effects were taken into account via GGA+$U$ approach as introduced in Ref.~[\citenum{Liechtenstein-95}]. We chose the on-site Coulomb repulsion parameter to be $U$=$8$~eV and $U$=$6$~eV for Cu and Fe respectively, while the Hund's rule coupling parameter ($J_H$) was taken as $J_H$=$0.95$~eV for both $3d$ metal transition ions~\cite{Ushakov-17,Pesant-11}. The occupation numbers of Fe-$3d$ states were obtained by integration within atomic sphere with radius $1.302$~\AA. The crystal structure in GGA+$U$ calculations was relaxed unless the interatomic forces were larger than 0.005 eV/\AA.

\section {Calculation results \label{DFT}}

We start with the simplest non-magnetic calculations, which results are presented in Fig.~\ref{gga:dos}. One might see that the bands extending from -2.7 to -1.5~eV correspond to Cu-$3d$ states, while Fe-$3d$ bands are between -1.5 to 2~eV. S-$3p$ bands are below -3~eV (not shown in  Fig.~\ref{gga:dos}a). There are 4 formula units in the unit cell used in the calculation. All twenty (twice degenerate due to spin) Cu-$d$ bands are below the Fermi energy and therefore we see that Cu must be 1+ ($3d^{10}$) in cubanite.
\begin{figure}[t]
\includegraphics[width=0.4\textwidth]{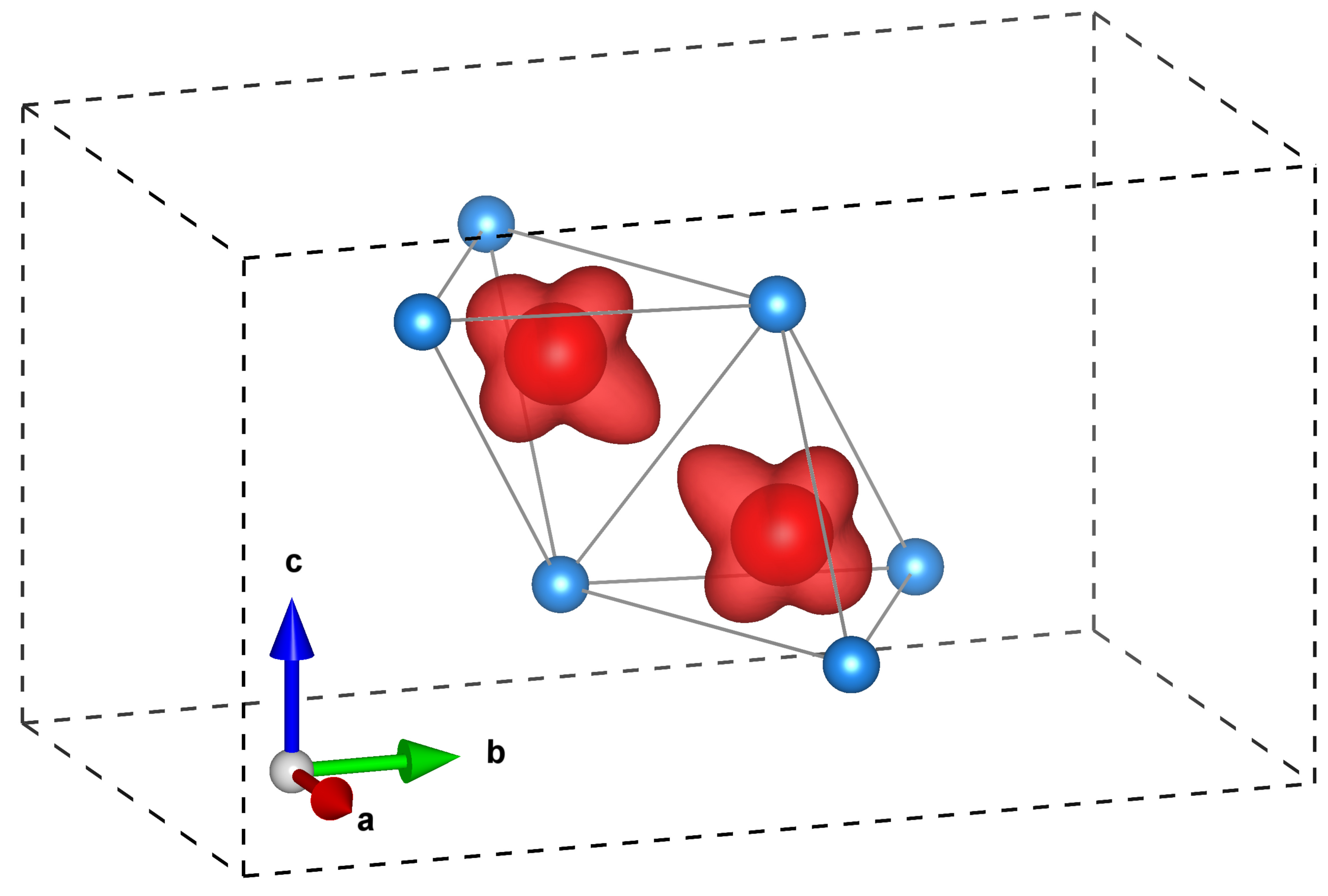}
\caption{\label{gga:par} The partial charge density represented 4 separated Fe-$3d$ bands on [$-1.1$, $-1.6$]~eV interval in a band structure on Fig.~\ref{gga:dos}(a).}
\end{figure}

The second important fact, which can be extracted from this type of calculations is that we see four lowest isolated Fe-$3d$ bands in energy range from -1.1 to -1.6 eV (Fig. \ref{gga:dos}a). Thus, there is one such band per each formula unit. A partial charge density corresponding to these bands shows that these are $e_g$ states ($x^2-y^2$ orbitals in the local coordinate system, where the axis points to the centers of tetrahedron edges), see Fig.~\ref{gga:par}. One can also see that these are exactly two $d$ orbitals directed to each other in the common edge geometry of two FeS$_4$ tetrahedra.

This explains M\"ossbauer data, which do not show charge disproportionation between two irons, but observe a ``rapid electron exchange''\cite{Imbert-67,Greenwood-68}, which is obviously a consequence of molecular orbitals formation. Moreover, one can see that there are two types of $d$ electrons in CuFe$_2$S$_3$ those forming molecular orbitals ($x^2-y^2$) and others, which are expected to be localized at atomic sites. Thus, cubanite could be an example of materials demonstrating orbital-selective behavior~\cite{Streltsov2014,Streltsov2016b}.

\begin{figure}[t]
\includegraphics[width=0.5\textwidth]{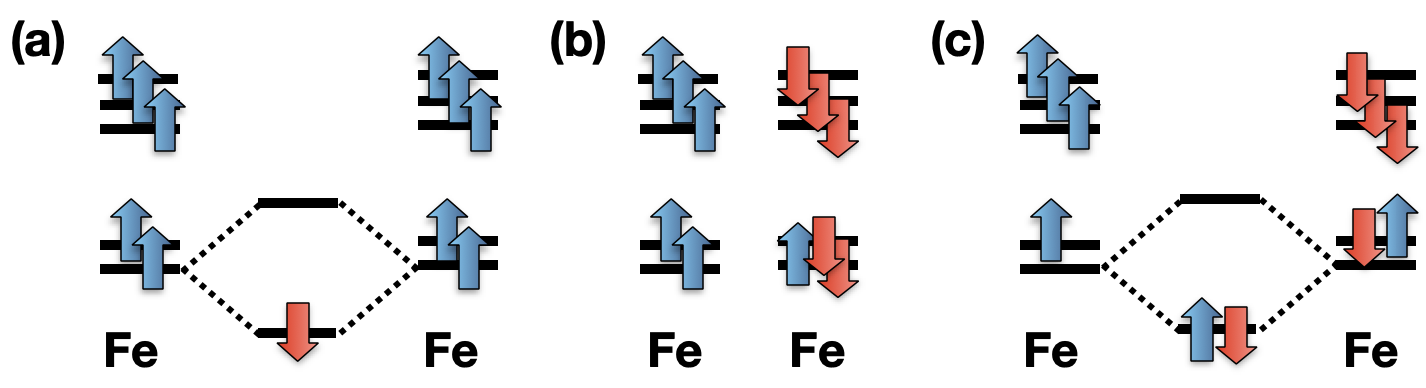}
\caption{\label{sketch} The sketch illustrating stabilization of the ferromagnetic order, case (a), due to the formation of molecular orbitals ($x^2-y^2$) in Fe--Fe pair. (b) and (c) demonstrate two other possible solutions with all electrons localized at sites (case b) and two electrons occupying the molecular orbital (case c). It is assumed that bonding-antibonding splitting is large and one can neglect superexchange between localized electrons.}
\end{figure}

There are important implications of orbital-selective physics in CuFe$_2$S$_3$. E.g. in the case of odd number of electrons on $d$ shell the mechanism like double exchange stabilizes ferromagnetic exchange interaction for nearest neighbor Fe ions (those forming dimers Fe$_2$S$_6$ with a common edge of FeS$_4$ tetrahedra). Neglecting conventional superexchange one can readily see that the ferromagnetic configuration shown in Fig.~\ref{sketch}a has the lowest total energy $E_{a} = -20 J_H - t$, where $t$ is a hopping parameter for electrons forming molecular orbitals and $J_H$ is the intra-atomic exchange. Antiferromagnetic order is impossible in this situation since one might have only one electron with given spin per site.  This state is always lower in energy than the configuration with site-localized electrons of  Fig.~\ref{sketch}b (only antiferro version is shown; the ferro one has the same energy if one neglects super-exchange), $E_b = - 20J_H$. There also will be the configurations with two electrons occupying the molecular orbital, one of them is shown in Fig.~\ref{sketch}c, it is energy is $E_c = - 16.5J_H - 2t$. Thus, we see that ferromagnetic ``intra-dimer'' order wins if $3.5 J_H > t$, which is exactly the case of  CuFe$_2$S$_3$, since $J_H \sim 1$ eV and $t= 0.4$~eV as follows from our DFT calculations (the value of $t$ was taken from Wannier function projection). 

This idea is very similar to a conventional double exchange: itinerant electrons with the $x^2-y^2$ symmetry move from one Fe site to another and thereby substantially lowers the total energy (Fig.~\ref{sketch}a). This is only possible if both Fe sites have the same spin projection. If there would be no such fluctuations, i.e. the system will be in configurations shown in Fig.~\ref{sketch}b or Fig.~\ref{sketch}c, there will be no such energy gain.

\begin{table}[b!]
\caption{\label{tab:mag} The total energies (E$_{tot}$) and energy gap (E$_{gap}$) for the insulating states as obtained from GGA and GGA+$U$ ($U$(Cu)=8 eV, $U$(Fe)=6 eV) calculations for the different magnetic structures of CuFe$_2$S$_3$ (per formula unit). Notation are the same as in Ref.~[\citenum{Townsend-73}], corresponding magnetic configurations are presented in Fig.~\ref{fig-mag}.}
\begin{ruledtabular}
\begin{tabular}{l|ll|ll}
 Magnetic & \multicolumn{2}{c|}{GGA} & \multicolumn{2}{c}{GGA+$U$} \\ 
 config. &  E$_{tot}$, meV & E$_{gap}$, eV & E$_{tot}$, meV & E$_{gap}$, eV \\
 \hline
 A & 0  & 0.3 & 0  & 0.8  \\
 C & 191.6 & 0.05 & 98.1  & 0.51  \\
AF dimers & 262.6  & no & 179.6  & no  \\
B & 621.2 & no & 335.4  & 0.35  \\
\end{tabular}
\end{ruledtabular}
\end{table}

This discussion is in accord with direct magnetic GGA calculations, which results are presented in Table~\ref{tab:mag}. The lowest in total energy is the configuration with Fe spins ordered ferromagnetically inside Fe--Fe dimers  and maximally antiferromagnetic between them (configuration A). This result agrees with analisys of magnetic data performed in Ref.~[\citenum{Townsend-73}]. It also has to be mentioned a conventional superexchange mechanism between localized electrons tends to stabilize AFM order and is indeed operative for inter-dimer exchange interaction, while intra-dimer ferromagnetism is due to the formation of molecular orbitals, i.e. orbital-selective behavior.


\begin{figure}[t!]
\includegraphics[width=0.4\textwidth]{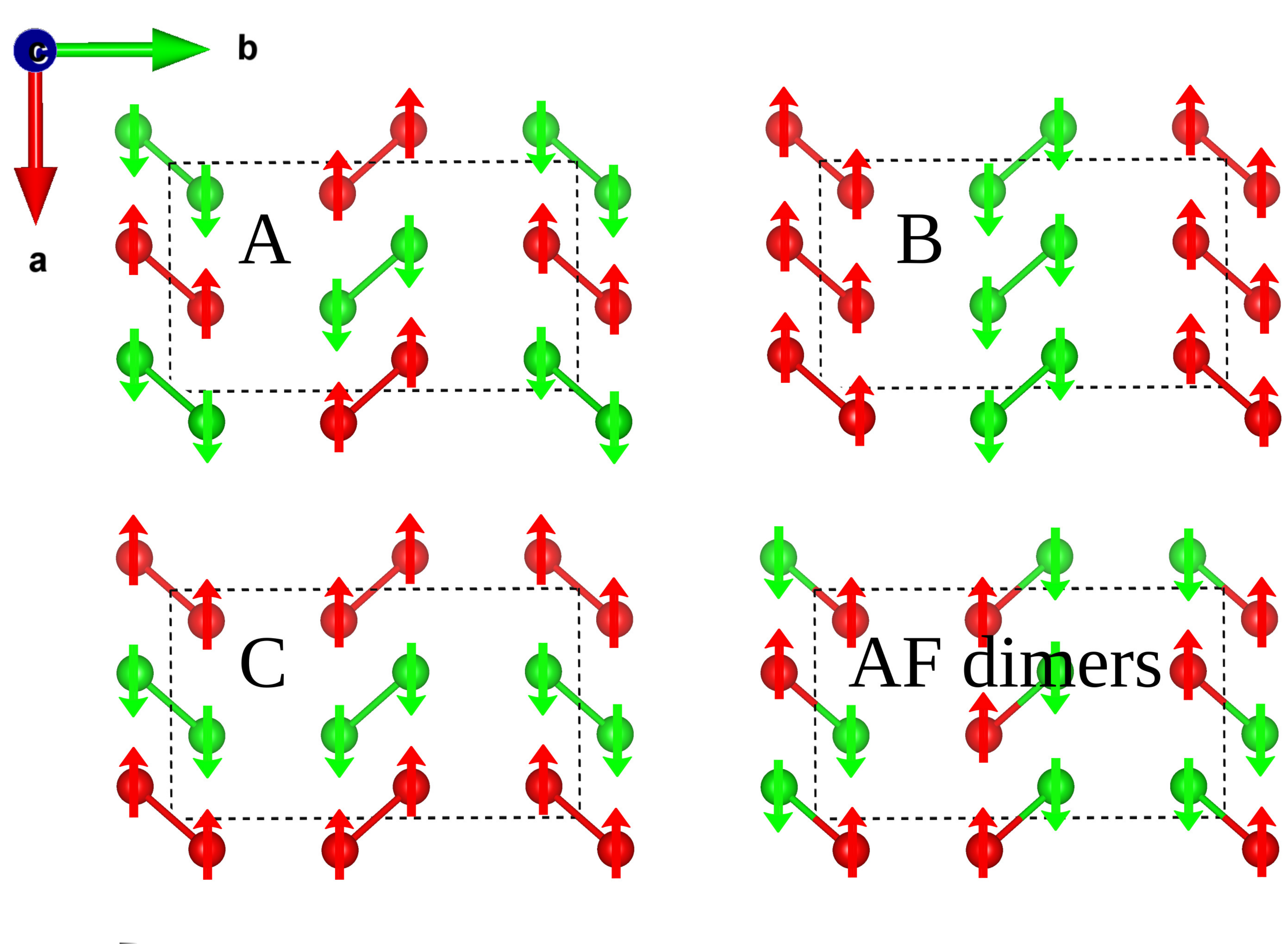}
\caption{\label{fig-mag} The schematic representation of the possible magnetic structures of CuFe$_2$S$_3$ discussed in Ref.[~\citenum{Townsend-73}] with FM order in Fe--Fe dimers and one more with AFM order in Fe--Fe dimers considered in this paper.}
\end{figure}

Moreover we see that our GGA calculations are able to reproduce the correct magnetic ground state, and CuFe$_2$S$_3$ becomes insulating already in GGA approximation. The energy gap in configuration A is about 0.3 eV. Thus the ferromagnetic order of Fe spin moments in Fe-Fe dimers retains an additional itinerant electron for two irons ions within a dimer cluster

\begin{figure}[t!]
\includegraphics[width=0.4\textwidth]{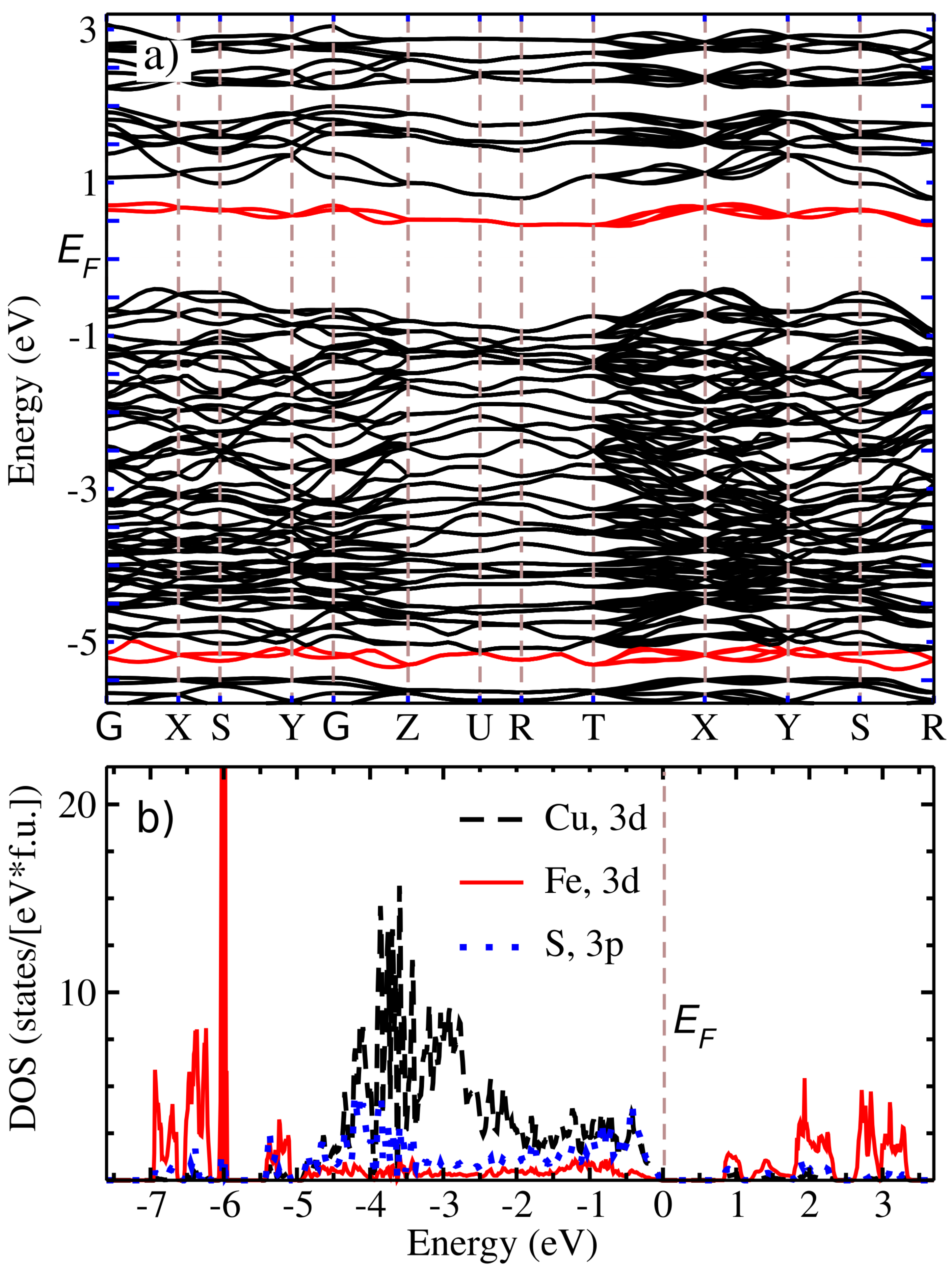}
\caption{\label{ggau:dos} The band structure (a) and partial densities of states (b) of CuFe$_2$S$_3$ obtained in GGA+U approximation ($U$(Cu) = $8$~eV, $U$(Fe) = $6$~eV, $J_H$(Cu, Fe) = $0.95$~eV). The bands with bonding-antibonding splitting of molecular orbitals are marked by red color, their character was deduced from the corresponding charge-density plots.}
\end{figure}

In order to take into account correlation effects in CuFe$_2$S$_3$, we carried out GGA+$U$ calculations. The results are shown in Fig.~\ref{ggau:dos} and displayed in Table~\ref{tab:mag}.
One can see that the band gap in GGA+$U$ increases up to about 0.8 eV in configuration A. The total energy calculations demonstrate that the state with FM dimer remains the lowest, see Table~\ref{tab:mag}. The decrease of the energy difference between various configurations is related to the fact that now not Stoner-like exchange, but a much larger Hubbard $U$ correction defines the superexchange processes (in the denominator for the superexchange interaction).

The detailed analysis of the occupation matrices shows that the charge disproportionation does not occur even if all symmetry information is removed in the calculations. This suggests that molecular orbitals are not destroyed completely by the Hubbard correction (which tends to localize all electrons on atomic orbitals), but a more accurate methods such as e.g. cluster DFT+DMFT calculations\cite{Biroli2002} should be used to study this effect. Moreover, lattice optimization also does not break the charge homogeneous state. Interestingly enough, we were able to obtain charge ordered solution in GGA+$U$ approach when Fe ions are coupled AFM in a Fe$_2$S$_6$ dimers, but its total energy is higher on $180$ meV.

 In order to justify the validity of chosen $U$ parameters we repeat the calculations using smaller $U$(Fe)=4.5 eV, as e.g. in Ref.\cite{Pchelkina2013}, and $U$(Cu)=8 eV ($J_H$ on both ions is 0.95 eV). The results remain the same: the magnetic configuration A corresponds to the ground state, which is insulating with a band gap is of $\sim$1 eV. We obtain for other configurations have total energies of: 725 (B), 125 (C), and 400 (AFM dimers) meV. The ground state also does not change if one ignores Hubbard $U$ on Cu (it also retains insulating behavior).

\section {Conclusions}

We used {\it ab initio} calculations to study the physical properties of CuFe$_2$S$_3$.  In transition metal compounds containing structural clusters of metals there can be realized a special state, when some of the electrons form singlet ($S=0$) pairs, while others are effectively decoupled and may give e.g. a long-range magnetic order or stay paramagnetic~\cite{Streltsov2014,Streltsov2016b}. Our GGA calculations show that there are indeed two types of $3d$-electrons in cubanite: those forming molecular orbitals ($x^2-y^2$) and others, which are expected to be localized at atomic sites. This fact determines CuFe$_2$S$_3$ as a system with orbital-selective behavior. This in turn strongly affects magnetic properties of cubanite making exchange interaction between nearest neighbors strong and ferromagnetic (by ``double-exchange-like'' mechanism), while magnetic ordering between other irons is governed by the superexchange path. 

The orbital-selective behavior also explains puzzling M\"ossbauer data, where no Fe separation onto Fe$^{2+}$ and Fe$^{3+}$, but a ``rapid electron exchange'' has been observed. We feel that our findings could be applied to explain the metallic high pressure phase of CuFe$_2$S$_3$, where Fe ions are surrounded by sulphur octahedra, forming dimers with a common face\cite{Rozenberg1997a}. In such a geometry a$_{1g}$ orbitals may take part in a strong molecular bonding leaving other electrons ($e_g^\pi$) site localized as it occurs in e.g. Ba$_5$AlIr$_2$O$_{11}$~\cite{Terzic2015,Streltsov2017} or hexagonal structure such as Ba$_3$Me$TM$O$_9$ (where $TM$ is a transition metal and Me is metal like Li, Na, La etc.)\cite{Nguyen2021}. Of course this has to be checked by corresponding calculations on the high pressure phase of CuFe$_2$S$_3$.

\section {Acknowledgments}
We acknowledge support of the Russian Science Foundation via project  20-62-46047 and also thank Prof. A. Fujimori for useful discussions on physical properties of this material.


\bibliography{apssamp}

\end{document}